\documentclass[%
 aip,
 amsmath,amssymb,
preprint,
]{revtex4-1}
\usepackage{graphicx}
\usepackage{dcolumn}
\usepackage{bm}

\usepackage[utf8]{inputenc}
\usepackage[T1]{fontenc}
\usepackage{mathptmx}
\usepackage{etoolbox}
\usepackage{subcaption}
\usepackage[colorlinks, linkcolor=blue,citecolor=blue, urlcolor=black]{hyperref}
\usepackage[sort&compress]{natbib}
\makeatletter
\def\@email#1#2{%
 \endgroup
 \patchcmd{\titleblock@produce}
  {\frontmatter@RRAPformat}
  {\frontmatter@RRAPformat{\produce@RRAP{*#1\href{mailto:#2}{#2}}}\frontmatter@RRAPformat}
  {}{}
}%
\makeatother
\begin{document}

\title[Global properties and stability  of transonic plasma acceleration in the magnetic nozzle]{Global properties and stability  of transonic plasma acceleration in the magnetic nozzle}
\author{N. Sheth}
\email{nishka.sheth@usask.ca.}
\author{A. Smolyakov}%
\author{J. Deguire}
\affiliation{Department of Physics and Engineering Physics, University of Saskatchewan, Saskatoon SK S7N 5E2, Canada.}
\author{\firstname{S.}~\surname{Pande}}
\affiliation{Indian Institute of Technology (BHU), Varanasi, India } 
\author{\firstname{P.N.}~\surname{Yushmanov}}
\affiliation{TAE Technologies Inc., 19631 Pauling, Foothill Ranch, California 92610, USA}  
\date{\today}

\begin{abstract}
It is shown that transonic plasma acceleration in the converging-diverging magnetic field (magnetic nozzle) follows the unique global solution which is fully defined by the magnetic field. Such solution, which was analytically obtained earlier in the paraxial approximation, is compared here with results of the axisymmetric two-dimensional ($r-z$) magnetohydrodynamics(MHD) simulations. It is shown that analytical solution describes well the region near the axis but also can be applied to arbitrary magnetic surfaces. The simulations with different length of the nozzle and different boundary values for plasma velocity show that the plasma flow switches to the unique transonic acceleration profile via the shock-like transition in the velocity and pressure profiles. The simulations with arbitrary (not vacuum) initial magnetic field demonstrate the global adjustment of the magnetic field such that the transonic acceleration velocity profile follows the analytic predictions with the modified magnetic field.
\end{abstract}

\maketitle

Expanding magnetic field (magnetic nozzle) and converging-diverging (magnetic mirrors) configurations are of interest for electric propulsion\cite{ArefievPoP2004,WuPoP2025,MerinoPoP2016}, material processing and advanced manufacturing\cite{SchoenbergPoP1998}, and fusion\cite{RyutovAIP2016,EndrizziJPP2023} applications. 
Magnetic mirrors,     
with converging-diverging magnetic field configurations are used to confine plasma in fusion devices. 

Plasma acceleration in the magnetic nozzle was demonstrated in early experiments\cite{AndersenPF1969,KurikiPF1970}. Transonic plasma acceleration guided by the magnetic field is a basis of electrodeless plasma thrusters for space propulsion\cite{BathgatePSST2017,WuPoP2025} and also considered in  plasma sources for advanced material processing\cite{SchoenbergPoP1998}.
Plasma expansion in the diverging magnetic field is a crucial  process in a divertor of  the  open mirror systems\cite{RyutovAIP2016,OnofriPoP2017}, which are used to accept and direct large energy fluxes to the material walls.  Plasma flow guided and controlled by the magnetic field is a key question in all these applications.  

General equilibrium equations for steady-state flows  were formulated    within one- and two-fluid magnetohydrodynamics(MHD)\cite{Morozov_V8,SteinhauerPoP1999}, in particular, for applications to the problem of equilibrium and stability of fusion plasmas\cite{HameiriPF1983,GuazzottoPoP2005}. Similar physics is involved in the problem of plasma flows in space\cite{GoedbloedPhysScr2002}, solar wind acceleration\cite{VelliASC2001} and astrophysical winds and jets\cite{Tsinganos2007}.
In general, these models generalize the Grad-Shafranov equilibrium problem to include the inertial forces into the equilibrium. 

\begin{figure}[!h]
    \centering
    \includegraphics[width=0.5\linewidth]{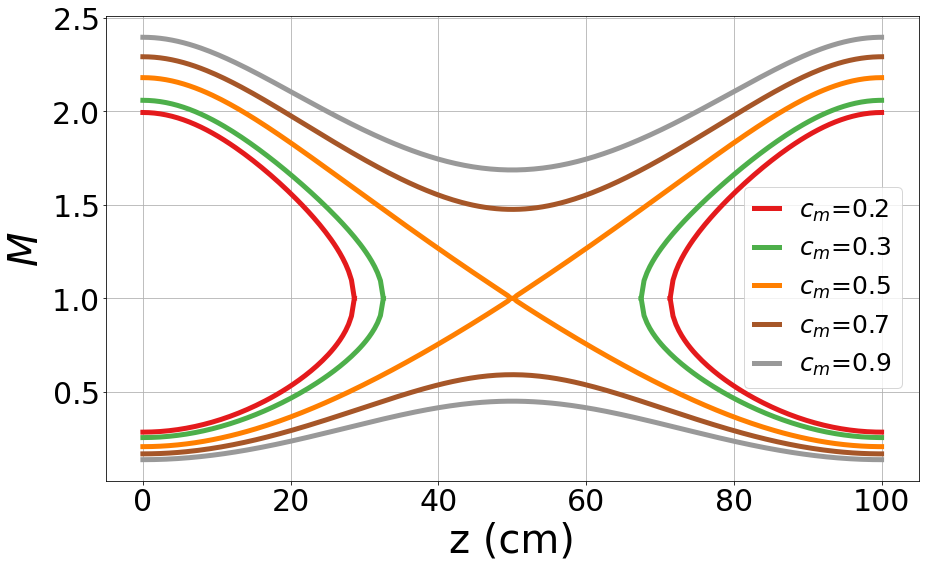}
    \caption{A typical diagram of the solutions for plasma flow in converging-diverging magnetic nozzle given by Eq.(\ref{eq:Mm}). The $c_m<0.5$ curves define the subsonic profiles, $c_m>0.5$ define supersonic profiles. The accelerating transonic profile is given by the separatrix with  $c_m=0.5$.}
    \label{fig:diagram}
\end{figure}

The balance of the inertial force with the electric field (or pressure) force is the main element of the acceleration and equilibrium in the magnetic nozzle. 
In the context of propulsion applications,  a simple model of transonic plasma acceleration can be formulated in the quasi-2D (paraxial) approximation, which can be reduced effectively to the 1D problem\cite{FruchtmanPoP2012, LafleurPoP2014,ManheimerIEEE2001,SmolyakovPoP2021}. For cold ions and 
isothermal electrons with $T_e=const$, the expression for the normalized ion velocity $M=V_\Vert/C_s$ can be reduced to the equation,
\begin{equation}\label{eq:Mm}
    \frac{M^2}{2}=\ln \left( \exp(c_m)M\frac{B_m}{B(z)} \right),   
\end{equation}
where, $C_s$ is ion-sound speed, $B=B(z)$ is the magnetic field as a function of the distance along the nozzle (near the axis ($r/a\ll 1$), and $B_m$ is the maximum value of the magnetic field. The value of the $c_m$ constant defines the particular solution. An example of the solutions diagram for the converging-diverging magnetic field used in this paper is shown in Fig.\ref{fig:diagram}. 
An exact solution of Eq.(\ref{eq:Mm}) can be written\cite{SmolyakovPoP2021} in terms of the Lambert function $W(y)$ defined as the solution of the equation $W\exp (W)=y$.
\begin{equation}\label{eq:Mz_theory}
    M(z) = \left[ -W\left[-b^2(z)\exp({-2c_m)}\right] \right]^{1/2}  
\end{equation}
where, $b(z) = B(z)/B^m$.
 
The plasma velocity along the transonic solution with $c_m=0.5$ is fully defined by the regularization condition at the magnetic throat and follows the magnetic field profile as expressed by Eq.(\ref{eq:Mz_theory}). This is a single unique transonic accelerating solution represented by the separatrix in Fig.\ref{eq:Mm}. The other separatrix, representing a decelerating solution, may be of interest for accreting objects in astrophysics or potentially for direct recuperation of the kinetic plasma energy into electricity, but is not considered here. The subsonic part, $M<1$, of the transonic accelerating solution is described by $M=[-W_0(-b^2(z)/e)]^{1/2}$ branch of the Lambert function, and the supersonic is given by the other branch $M=[-W_{-1}(-b^2(z)/e)]^{1/2}$. The full accelerating solution is formed by two branches $W_0(y)$ and $W_{-1}(y)$ which match smoothly\cite{DubinovJPP2005,CorlessAdvCompMath1996} at the point $M=1$ corresponding to the maximum magnetic field $B_m$.

For the converging-diverging magnetic nozzle with the maximum magnetic field $B_m$ at the throat, all possible solutions in Fig.\ref{eq:Mm} can be separated in several groups and can be classified by the value of the velocity at the nozzle inlet, $z=0$.
At the nozzle inlet, the transonic solution has $M(0)=[-W_0(-(B(0)/B_m)^2/e)]^{1/2}$, and at the nozzle end $M(L)=[-W_{-1}(-(B(L)/B_m)^2/e)]^{1/2}$. The solutions with $V_\Vert/C_s<M(0)$ remain purely subsonic, and for $V_\Vert/C_s > [-W_{-1}(-(B(0)/B_m)^2/e)]^{1/2}$ remain purely supersonic in the whole region. The solutions with the boundary values (at $z=0$) between two separatrices $[-W_{0}(-(B(0)/B_m)^2/e)]^{1/2}<V_\Vert/C_s< [-W_{-1}(-(B(0)/B_m)^2/e)]^{1/2}$  are multi-valued and not physical. 
The analytical solution in Eq.(\ref{eq:Mz_theory}) was obtained in the simplified (effectively 1D) case corresponding to the paraxial limit. It was found that such solutions are robust as they were realized in drift-kinetic and fluid simulations within similar 1D paraxial models, i.e. plasma flow converges to the analytical transonic solutions for a rather wide range of initial conditions\cite{JimenezPoP2022,SaboPoP2022,TyushevPoP2025}. It is relatively straightforward to show theoretically that the model in Ref.\onlinecite{SmolyakovPoP2021} can be reformulated in terms of the 1D problem along the arbitrary magnetic field surfaces. The goal of this article is to investigate plasma acceleration in full 2D axisymmetric ($r-z$) geometry and compare it with the analytical predictions. We demonstrate here that the theoretical solution in Eq.(\ref{eq:Mz_theory}) remains valid when applied along the magnetic surfaces. The robust behavior of the transonic acceleration in 2D MHD model was shown preliminary in Ref.\onlinecite{DeguireMSc}. 

We employ here the standard ideal one-fluid MHD. One can show that in the stationary state the axisymmetric plasma flow in one-fluid MHD exactly follows the magnetic field surface\cite{Morozov_V8,HameiriPF1983}. This condition exactly matches the approximations of the paraxial model in Ref.\onlinecite{SmolyakovPoP2021}. It is important to note that most of the earlier models for plasma flow in the magnetic nozzle are based on the assumption that the underlying magnetic field (from the external coil) is fixed and does not change. Here, we relax this assumption and allow the full modification of the magnetic field which is in the final saturated state determined self-consistently within the MHD evolution equations.    

The Standard equations of the ideal one-fluid MHD in the conservative form are written as follows, 
\begin{subequations}\label{eq:ideal_MHD}
\begin{align}
\frac{\partial \rho}{\partial t} + \nabla \cdot (\rho \mathbf{V}) &= 0, \\
\frac{\partial (\rho \mathbf{V})}{\partial t} + \nabla \cdot \left(\rho \mathbf{V} \mathbf{V} - \mathbf{B} \mathbf{B} + p_T \mathbf{I} \right) &= 0, \\
\frac{\partial \epsilon_T}{\partial t} + \nabla \cdot \left[\mathbf{V}(\epsilon_T + p_T) - \mathbf{B}(\mathbf{V} \cdot \mathbf{B}) \right] &= 0, \\
\frac{\partial \mathbf{B}}{\partial t} + \nabla \cdot (\mathbf{V} \mathbf{B} - \mathbf{B} \mathbf{V}) &= 0.
\end{align}
\end{subequations}
Here, the total plasma pressure $p_T = p + \frac{B^2}{2}$, and the total internal energy density $\epsilon_T = \frac{1}{2} \rho V^2 + \rho \epsilon + \frac{B^2}{2}$ . The standard notations are used throughout the paper for the physical variables: mass density, plasma flow velocity, magnetic field, kinetic pressure, and internal energy as $\rho$, $\mathbf{V}$, $\mathbf{B}$, $p$, and $\epsilon =p/\rho(\gamma -1)$ respectively. 

In our base case, we initialize simulations from a vacuum force-free magnetic field $\mathbf{B}_0$ defined by the poloidal flux function $\psi$ in the form\cite{PostNF1987}, 
\begin{equation}\label{eq:B_psi}
\mathbf{B}_0 = \frac{1}{r} \left( -\frac{\partial \psi}{\partial z}, 0, \frac{\partial \psi}{\partial r} \right),
\end{equation}
with,
\begin{equation}\label{eq:psi}
\psi =  \frac{r^2 B_*}{2}  \left(1 - \frac{\alpha L}{ \pi r}\cos\left(\frac{2 \pi  z}{L}\right) I_1\left(\frac{2 \pi r}{L}\right)\right),
\end{equation}
giving the magnetic field in the form, 
\begin{align}
B_{r}&=-B_* \alpha \sin \left(\frac{2 \pi z}{L}\right) I_{1}\left(\frac{2 \pi r}{L}\right), \label{eq:Br_vacume}\\
B_{z}&=B_*\left(1-\alpha \cos \left(\frac{2 \pi z}{L}\right) I_{0}\left(\frac{2 \pi r}{L}\right)\right).\label{eq:Bz_vacume}
\end{align}
For the symmetric nozzle,  $0<z<L$, the resulting mirror ratio $R$,  and the expansion ratio $K$ are equal $ R = K = (1 + \alpha)/(1 - \alpha)$, where $R=B_m/B(0)$ and $K=B_m/B(L)$, and $B_m=B_*(1+\alpha)$. We have taken here the value $\alpha=0.5$.

The ideal MHD Eqs.\eqref{eq:ideal_MHD} are solved in the 2D cylindrical $r-z$ geometry using the Godunov-type open source MHD code \textsc{Pluto} v4.3\cite{Mignone2007}. The cylindrical nozzle domain of radius $a=20$ cm and the axial length $L=100$ cm, is uniformly divided into $156 \times 256$ grid points. We chose the van Leer slop limiter, \textsc{HLL} Riemann solver, parabolic reconstruction scheme, and \textsc{RK}3 time-stepping with the Courant number 0.5. 

In general, the magnetic field is evolving in time. For the base case, with the initial force-free field, the total magnetic field $\mathbf{B}$ is decomposed into a time-independent background component $\mathbf{B_0}$, given by Eq.(\ref{eq:B_psi}) and a perturbation $\mathbf{B_1}$, such that $\mathbf{B}=\mathbf{B_0}+\mathbf{B_1}$. The magnetic field in the form of Eq.(\ref{eq:B_psi}) inherently satisfies the solenoidal condition and is curl-free. The background magnetic field, $\mathbf{{B}_0}$, is held constant at all boundaries, while the perturbations of $\mathbf{{B}_1}$ remain free to evolve. This enables the Background Splitting option in the PLUTO code. The divergence-free condition on the $\mathbf{B_1}$ is maintained through the application of the constrained transport algorithm.  

We initialize computations from the stationary plasma state with $\mathbf{V}=0$ and uniform density $\rho_0$. At the axis of symmetry, $r = 0$, an axisymmetric condition is enforced;  the outer radial boundary $r = a$ fixes the $\rho$ and $\mathbf{V}$ to their initial values. At the nozzle inlet, $z=0$, the values of plasma density and axial velocity ($V_z$) are  maintained as  boundary conditions:     
     \begin{align}
          \rho &=4 m_H n_0 {\text{sech}}^2\left(\kappa^2 \left(\frac{r}{a}\right)^2 \right) + \rho_0, \\
         V_z &= v_0 C_s {\text{sech}}^2\left(\kappa^2 \left(\frac{r}{a}\right)^2 \right) ,    
    \end{align}
providing localized plasma injection with the characteristic radius $a/\kappa$, where we take $\kappa= 2.6$. At the exit of the nozzle, $z = L$, the outflow condition in imposed. 

Here, sound speed $C_s=\gamma p/\rho$. For the density we use $\rho_0 = 0.2 m_H n_0 \,\text{g/cm}^3$, with  $n_0 = 10^{12}\,\text{cm}^{-3}$, $m_H$ is Hydrogen mass. In general, in various fusion and propulsion scenarios, the plasma pressure can be anisotropic. It is often characterized by some effective value of $\gamma$. Effects of varied $\gamma$ on the acceleration were studied in Ref.\onlinecite{SmolyakovPoP2021}. Since it is not essential for the results of this paper, and for  simplicity, we use $\gamma=1$, so that $C_s=\sqrt{k_BT/m_H}$ for chosen temperature $T= 300$eV. In this paper, we express time in the units of Alfven time, denoted as $\tau_A = L/V_A$. Here, $V_A$ refers to the Alfven velocity, which is defined as $V_A = B_m/ \sqrt{4 \pi m_H n_0}$. Additionally, in what follows, the density is normalized by $\rho_m = 4 m_H n_0 + \rho_0$.

\begin{figure}[!h]
\begin{minipage}{0.49\linewidth}
        \centering
        \includegraphics[width=\linewidth]{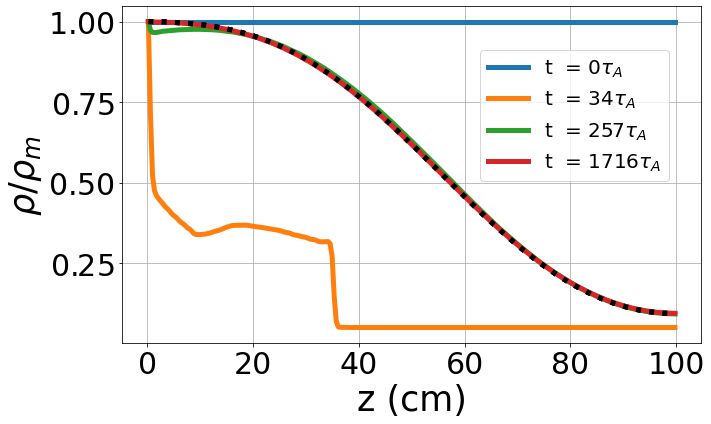}
        \subcaption{}
        \label{fig:rho_diff_time}      
    \end{minipage}
    \hfill
    \begin{minipage}{0.49\linewidth}
        \centering
        \includegraphics[width=\linewidth]{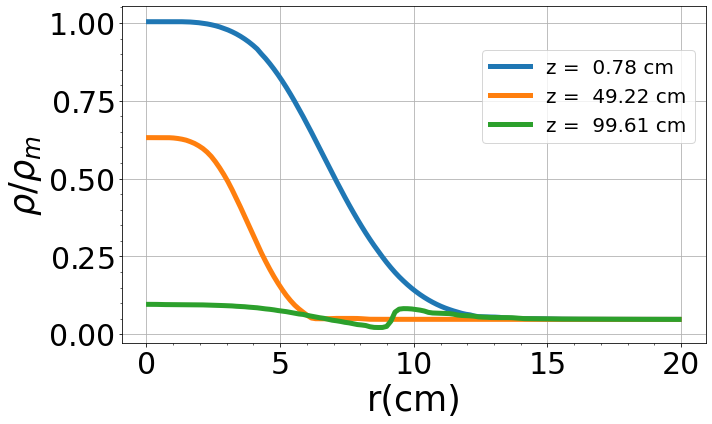}
        \subcaption{}
        \label{fig:rho_theroy} 
    \end{minipage}
    \caption{(a) Time evolution of the plasma density toward the stationary state. The axial profiles of the density are shown at different times as labeled. The theoretical profile from the Eq.(\ref{eq:Mz_theory}) and mass conservation is  shown by the dotted line. (b) The radial  profiles of the density in the stationary state are shown at different locations in $z$ as labeled. } 
\end{figure}

\begin{figure}[!h]
\begin{minipage}{0.49\linewidth}
        \centering
        \includegraphics[width=0.8\linewidth]{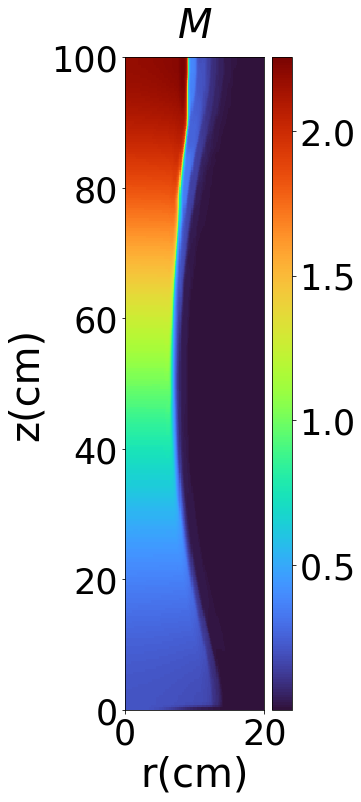}
        \subcaption{}
        \label{fig:pseudo_plot_v}      
    \end{minipage}
    \hfill
    \begin{minipage}{0.49\linewidth}
        \centering
        \includegraphics[width=0.8\linewidth]{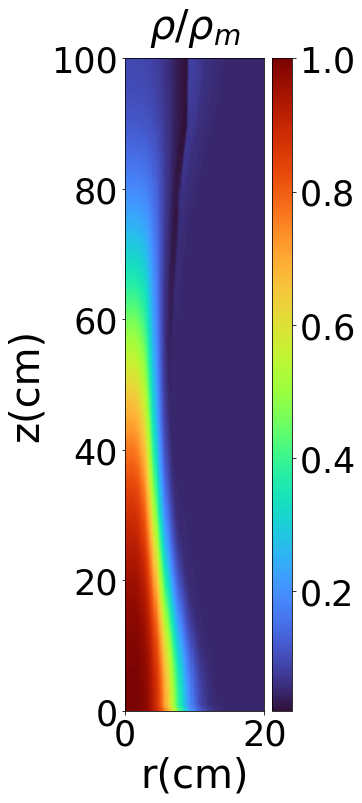}
        \subcaption{}
        \label{fig:pseudo_plot_rho} 
    \end{minipage}
    \caption{Two-dimensional profiles of (a) plasma velocity and (b) density demonstrating the transonic acceleration. The sonic point transition, $M=1$ occurs at the magnetic throat (maximum magnetic field) located at $z=50$ cm.}   \label{fig:2Dprofiles}
\end{figure}

The stationary flow equilibrium is established over the characteristic ion sound time $\tau_s=L/C_s$ which is much longer compared to the Alfven time due to low plasma pressure $\beta=\tau_s^2/\tau_A^2\ll 1$ for our parameters. The time evolution of the axial plasma density toward the equilibrium is shown in Fig.\ref{fig:rho_diff_time}, and the final stationary state is established after several hundred $\tau_A$. The radial density distribution essentially follows the boundary condition profile as demonstrated by Fig.\ref{fig:rho_theroy}, which shows the radial density profile at the different $z$ locations.  The 2D profiles of the plasma velocity and density in the stationary state, as shown in Figs.\ref{fig:pseudo_plot_v} and \ref{fig:pseudo_plot_rho},  demonstrate supersonic acceleration with the sonic point $V_\Vert=C_s$ at the location of the maximum magnetic field (nozzle throat). The time evolution of these profiles, highlighting the dynamic transition to the stationary state, is shown in the supplementary video.    

\begin{figure}[!h]
    \centering
    \includegraphics[width=0.7\linewidth]{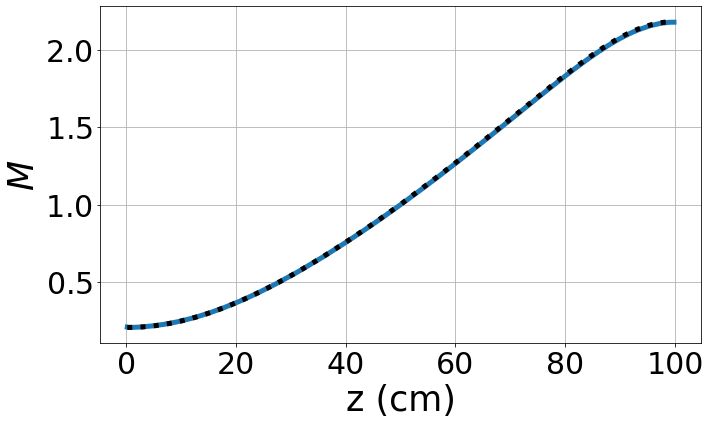}
    \caption{The comparison of the stationary velocity profile obtained in simulations as a function of $z$, for   $r = 0$ and  $t= 1716 \tau_A$ , which shown by the solid line, with the  analytical solution given by Eq.(\ref{eq:Mz_theory}), shown by the dotted line.}
    \label{fig:Mz_theory}
\end{figure}

We find an excellent agreement of the actual velocity profile (as a function of $z$ near the axis $r=0$) obtained in simulations with the analytical result of Eq.(\ref{eq:Mz_theory}) as shown in Fig.\ref{fig:Mz_theory}. For this comparison we have used the magnetic field as determined by Eqs.\eqref{eq:B_psi}-\eqref{eq:Bz_vacume}. 

\begin{figure}[!h]
    \centering
    \includegraphics[width=0.8\linewidth]{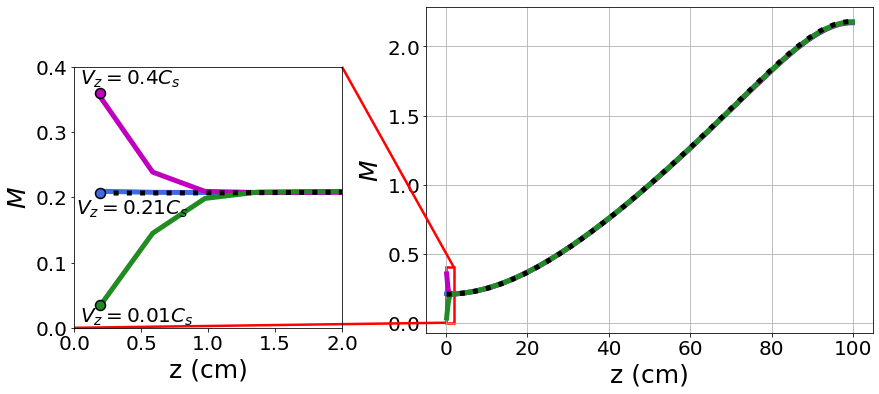}
   \caption{Transonic acceleration velocity profiles, $M=V_\Vert/C_s, $ for different boundary values of $V_z$ at $z=0$, as a function of $z$; simulations -- solid lines, Eq.(\ref{eq:Mz_theory})- dotted line.  The unique value of the velocity $V_z=0.21 C_s$ as predicted by Eq.~(\ref{eq:Mz_theory}) at $z=0$. The insert shows the behavior near $z=0$, where the transition to the unique profile occurs over few grid cells.} 
    \label{fig:robust}
\end{figure}

Our simulations demonstrate a remarkable stability of the transonic acceleration profile with respect of the variations in the boundary conditions. This is shown in the insert of Fig.\ref{fig:robust}, the stationary solution quickly transits into the unique transonic profile defined by Eq.\eqref{eq:Mz_theory}, even when the boundary value of the axial velocity at the inlet $z = 0$ is arbitrarily changed from its unique value of $V_z=0.21 C_s$ found from Eq.\eqref{eq:Mz_theory} for a given magnetic field. 

\begin{figure}[!h]
\begin{minipage}{0.49\linewidth}
        \centering
        \includegraphics[width=\linewidth]{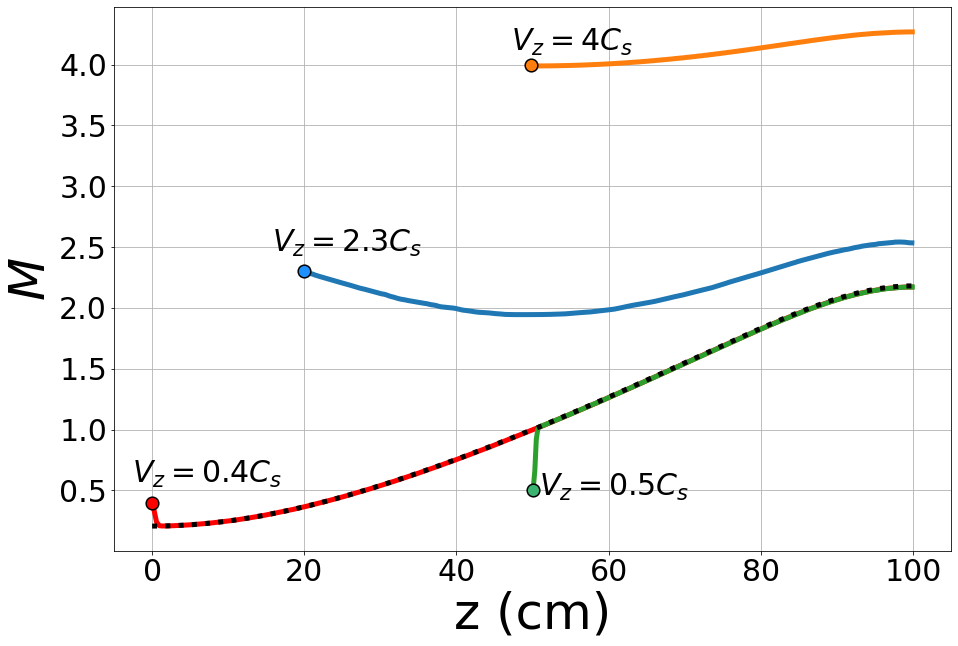}
        \subcaption{}
        \label{fig:M_z_space}      
    \end{minipage}
    \hfill
    \begin{minipage}{0.49\linewidth}
        \centering
        \includegraphics[width=\linewidth]{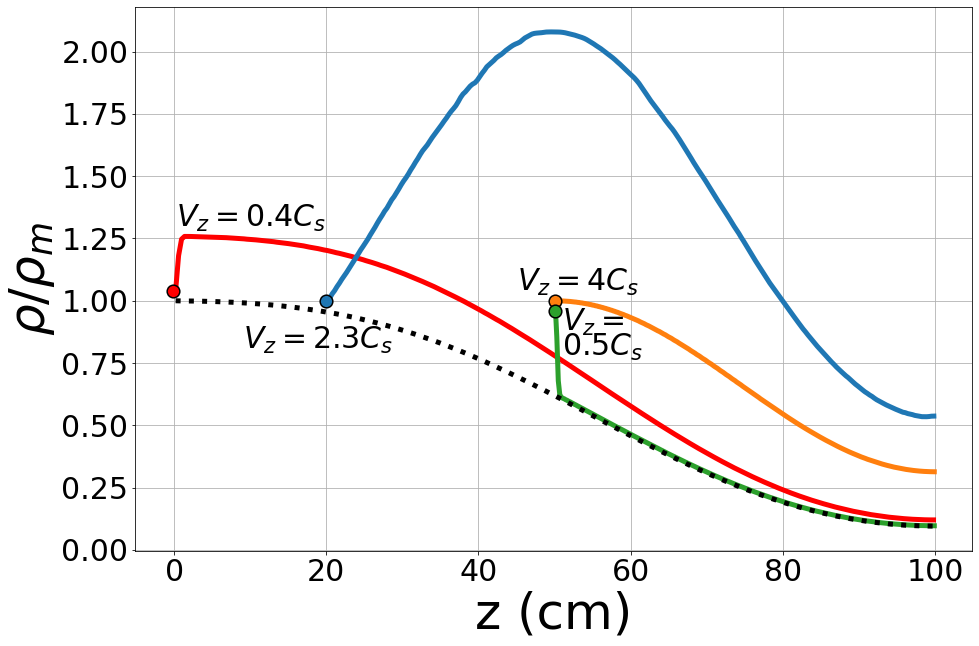}
        \subcaption{}
        \label{fig:rho_space}           
    \end{minipage}
    \caption{Comparison of the theoretical and numerical results for (a) plasma velocity and (b) density for different nozzle lengths and different injection velocities.  The black dotted lines are  from the theoretical solution from Eq.(\ref{eq:Mz_theory}) for  the 0--100 cm domain. The solid lines are the numerical results:  red -- for the 0-100 cm domain; blue for 20-100 cm domain; green and orange -50-100 cm domain.  The boundary values of the injection velocity at the left are as labeled; the boundaries  values for the pressure are the same in all cases.}
    \label{Combained fig: rho and M_z space}
\end{figure}

This property is further illustrated by the results in Fig.\ref{fig:M_z_space}, which shows several examples of the injection into the nozzle with different lengths and with different boundary values of the axial velocity. Thus, plasma injection with the boundary values of the velocity in the forbidden zone, between two separatricies in Fig.\ref{fig:diagram}, with the $V_z=0.4C_s$ also shown in Fig.\ref{fig:robust}, results in the transonic acceleration.  We recall that in the supersonic (and subsonic) zone there are multiple supersonic (and subsonic) solutions with different  boundary values. However, the injection with the velocity in the subsonic zone does not give a fully subsonic flow. Such boundary condition also results in  the transonic acceleration, as shown in Fig.\ref{fig:robust} for the nozzle 0-100 cm and $V_z=0.01C_s$ at $z=0$, and in Fig.\ref{fig:M_z_space}  for the nozzle  50-100 cm and with $V_z=0.5C_s$ at $z=50$ cm. This suggests that the subsonic solutions are unstable. On the contrary, the supersonic solutions are stable as it is shown in Fig.\ref{fig:M_z_space} for two examples: for the 20-100 cm nozzle with the boundary  value of $V_z=2.3 C_s$ at $z=20$ cm, and for the expanding nozzle of 50-100 cm and the injection value of $V_z=4C_s$ at $z=50$ cm. Fig.\ref{fig:rho_space} shows the behavior of the pressure profiles which are consistent with the corresponding cases in Fig.\ref{fig:M_z_space}. 

\begin{figure}
    \centering
    \includegraphics[width=0.5\linewidth]{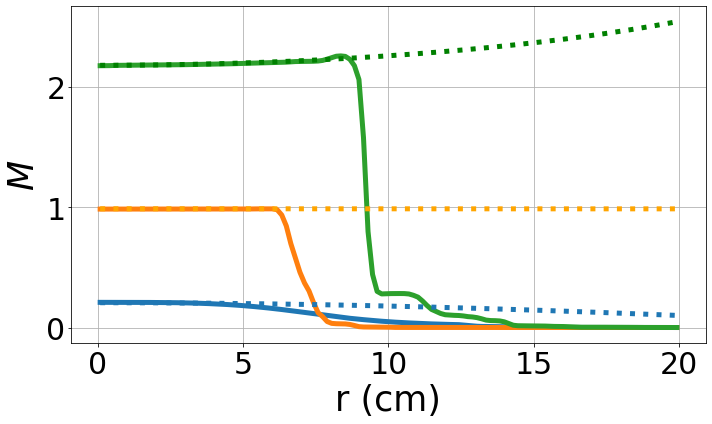}
    \caption{$M$ as a function of the  radius at  $z = 0$, $50$, \text{and} 
 $100$ cm, shown by the blue, orange, and green lines, respectively. The solid lines represent the simulation results, and the dotted lines correspond to the theoretical calculations from Eq.~(\ref{eq:Mz_theory}).}
    \label{fig:Mz_surface}
\end{figure}

The above examples show that the unique transonic profile given by the analytical solution is robustly valid near the axis $r=0$, i.e. in the paraxial approximation as it was initially derived in Ref.[\onlinecite{SmolyakovPoP2021}].  In this paper we  show furthermore  that the unique transonic solution remains valid when applied along the arbitrary magnetic surface. Thus, the Fig.\ref{fig:Mz_surface} shows the stationary profiles of the velocity as a function of the radius at three different locations in $z=0$, $z=50$ cm, and $z=100$ cm.  The dotted lines show the values of the theoretical plasma velocity calculated from Eq.(\ref{eq:Mz_theory}) for each location showing excellent agreement with the velocities obtained
in simulations (solid lines) for finite values of $r$ far from the axis.

The above examples used the initial magnetic field from Eq.(\ref{eq:psi}), which was a vacuum force-free field. In these initial value simulations, the magnetic field evolves self-consistently, thus allowing changes in the magnetic field  according to the full system of MHD Eqs.\eqref{eq:ideal_MHD}.  We find that in the case of the initial vacuum field, the modification of the magnetic field case was small but could not be neglected in the momentum balance.  To further explore the properties of the unique transonic acceleration, we have also performed initial value simulations starting  from different magnetic field given by the following model, 
\begin{equation}\label{eq:coil}
    {\psi} =\frac{r^2}{2}\frac{B_m}{[1+(R-1)((z-z_0)/L)^{2}]^{3/2}}.
\end{equation}
The model flux function corresponds to the paraxial limit, and satisfies the condition $\nabla \cdot \mathbf{B}=0$ to the first order in the radial small parameter $r/a\ll 1$. However, the field from   Eq.(\ref{eq:coil}) is not the current-free vacuum field,  $J_\phi \neq 0$.  



 

In this case, one cannot use the background field option of PLUTO, but the other simulation parameters and settings, including the initial and boundary conditions, remain the same as in the base case with the field from Eq.(\ref{eq:psi}).  The boundary conditions for the magnetic field at the radial boundary $r=a$ were imposed by fixing the values in the guard cells to the value of the initial magnetic field. 

\begin{figure}[!h]
         \centering
        \includegraphics[width=0.6\linewidth]{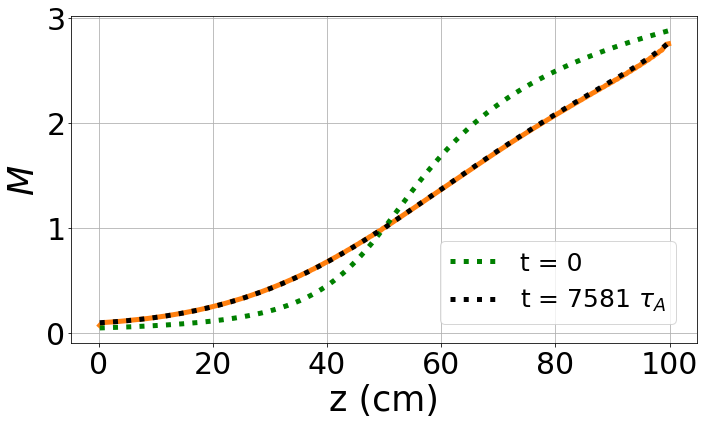}        
        \caption{Axial plasma velocity along the axis, $r=0$. Solid orange line- from simulations; dashed green line -  calculated using the Eq.(\ref{eq:Mz_theory}) with the original magnetic field from Eq.(\ref{eq:coil}), dashed black -calculated using the Eq.(\ref{eq:Mz_theory}) with the modified field from the simulations.}
        \label{fig:Mz_coil}
 \end{figure}

\begin{figure}[!h]
      \begin{minipage}{0.49\linewidth}
        \centering
        \includegraphics[width=\linewidth]{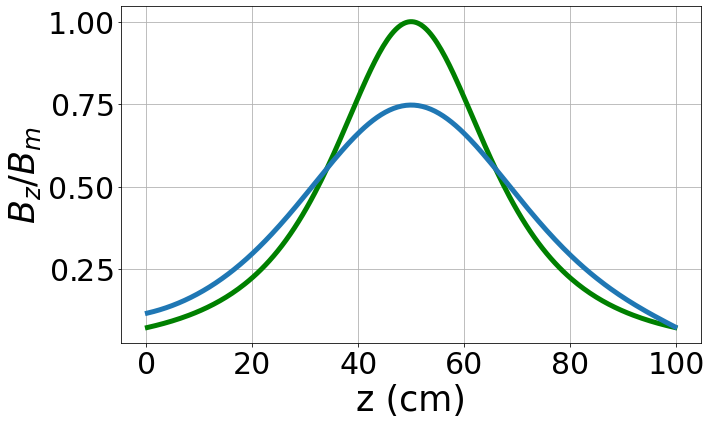}
        \subcaption{}
        \label{fig:Bnew_z}           
    \end{minipage}
    \begin{minipage}{0.49\linewidth}
        \centering
        \includegraphics[width=\linewidth]{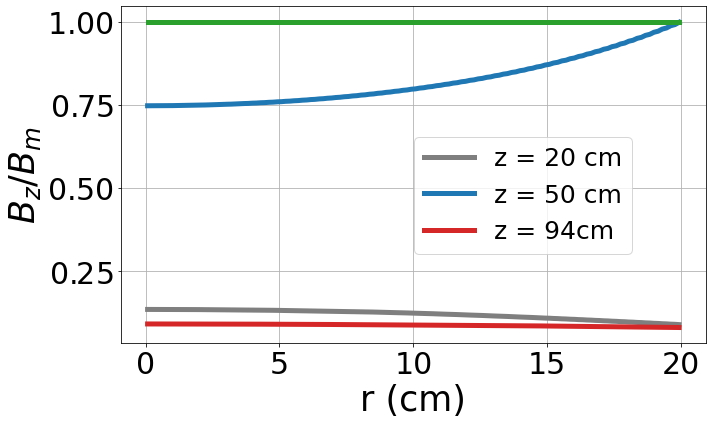}
        \subcaption{}
        \label{fig:Bnew_r}          
    \end{minipage}
     \caption{(a) The axial profile of the $B_z$  magnetic field: green line-original field from   from Eq.(\ref{eq:coil}), blue - the modified field from the simulations in the steady state. (b) The radial profile of the $B_z$ at different $z$ locations. The green line is the original magnetic field  line at $z=50$ cm, the blue line is the modified magnetic at $z=50$ cm. The red and grey lines show the modified $B_z$ field at the locations as labeled.}
    \label{fig:Bnew}
\end{figure}

\begin{figure}[!h]
    \begin{minipage}{0.49\linewidth}
        \centering
        \includegraphics[width=\linewidth]{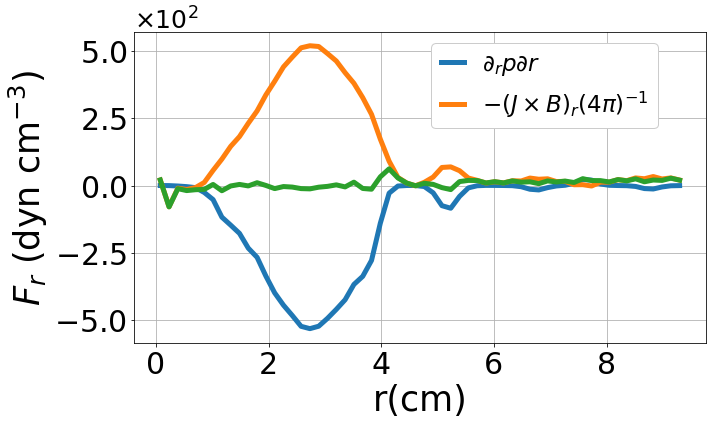}
        \subcaption{}
        \label{fig:force_r}
    \end{minipage}
    \hfill
    \begin{minipage}{0.49\linewidth}
        \centering
        \includegraphics[width=\linewidth]{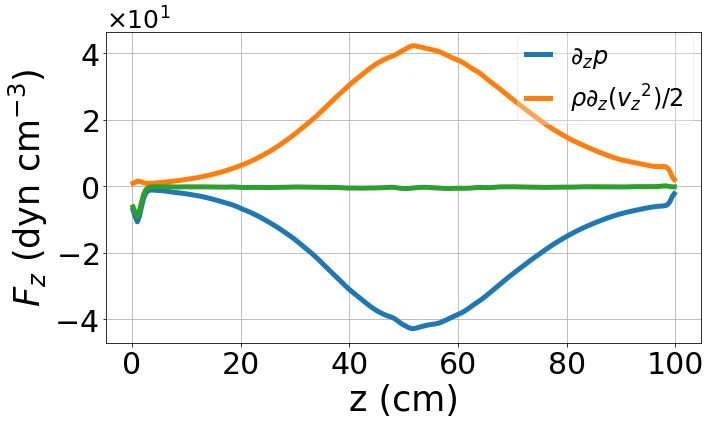}
        \subcaption{}
        \label{fig:force_z}
        \end{minipage}
     \caption{ The momentum balance for the main components in the axial (a) and radial (b) directions. (a) The sum of the  pressure gradient and Lorentz forces is shown by the green line. (b) The sum of the pressure gradient and the axial inertial forces is shown by the green line.  The other terms in the momentum balance are smaller at least by one order of magnitude  both for (a) and (b).}
    \label{fig:momentum}
\end{figure}

Similarly to the previous case we observe the formation of the stationary transonic acceleration. This solution is shown by the solid orange line in Fig.\ref{fig:Mz_coil}. However, the numerical solution does not agree with the analytical solution from Eq.(\ref{eq:Mz_theory}) and using the original magnetic field from Eq.(\ref{eq:coil}), shown by green dotted line and labeled as $t=0$. The discrepancy occurs because the initial magnetic field is modified in the course of the evolution. These modifications are shown in Figs.\ref{fig:Bnew_z} and \ref{fig:Bnew_r}. Fig.\ref{fig:Bnew_z} shows that compared to the initial field, the axial profile of the $B_z$ field in the stationary state at $t=7581 \tau_A$ becomes less peaked with a reduced amplitude at the maximum. The original magnetic field has a uniform profile radially, but the modified field is depressed in the radial direction toward the center; the value of $B_z$ at the boundary remains the same, as shown in  Fig.\ref{fig:Bnew_r}. The radial magnetic field has also changed. The numerical transonic profile, shown by the solid orange line in Fig.\ref{fig:Mz_coil} shows excellent agreement with the analytical solution along the modified field near $r=0$, as shown by the dotted line at $t=7581 \tau_A$. In the stationary state, the magnetic field is modified such that to satisfy the force balance as shown in Fig.\ref{fig:force_r} and Fig.\ref{fig:force_z}. Radially, the dominant terms are the radial pressure gradient balanced by the Lorentz force, Fig.\ref{fig:force_r}. In the axial direction, the axial pressure gradient force is balanced by the inertial force due to the acceleration, Fig.\ref{fig:force_z}.    

In this study, we extended the previous analytical results for the one-dimensional paraxial model to the full 2D ($r-z$) geometry of the magnetic nozzle by performing full one-fluid ideal MHD simulations. We show here that the analytical solution obtained in Ref.\onlinecite{SmolyakovPoP2021} remains valid when applied to the arbitrary magnetic surfaces such that the plasma is transonically accelerated along the magnetic surfaces according to the analytical predictions from Eq.(\ref{eq:Mz_theory}). Furthermore, we show here that within the full MHD model,  the underlying magnetic field guiding the plasma is self-consistently modified to satisfy the momentum balance. In the radial direction,   the primary balance is between the radial pressure gradient and the Lorentz force. In the axial direction, the primary balance describes the axial plasma acceleration by the axial pressure gradient force. In both directions, additional contributions from radial flows are small for our parameters.

It is important to note that the significant modification of the magnetic field observed in the case of a rather arbitrary magnetic field is not an effect of the finite plasma pressure. For our parameters, the plasma pressure parameter is rather small $\beta<1 \% $. In some sense, the ``self-healing'' of the magnetic field is an artifact of the arbitrary magnetic field in the initial state, which was not current free and therefore cannot represent the vacuum field. In presence of plasma, the MHD evolution (of the magnetic field, flow and plasma pressure) drives the system toward a new equilibrium  that (a) provides the thermal axial plasma acceleration in the magnetic nozzle conditioned by the converging-diverging magnetic field at the radial boundary; (b) satisfies the radial momentum balance which is primarily the balance of the radial pressure gradient and the Lorentz force; (c) plasma acceleration along the magnetic surface follows the earlier analytical predictions. We note that recent results\cite{PiochThesis2024} show that the plasma flow velocity measured in the plume of the magnetic nozzle of the Electron Cyclotron Thruster is indeed in agreement with the theoretical predictions of Ref.\onlinecite{SmolyakovPoP2021}.  We have to note that the effects of plasma rotation,  axial current, and the two-fluid (Hall) effects are not included in these simulations or the theoretical model\cite{SmolyakovPoP2021}. The effects of plasma rotation and azimuthal magnetic field (due to the axial current) on plasma acceleration were considered in one -fluid MHD model in Ref.\onlinecite{SmolyakovPPR2025}. 
\section*{Supplementary Material}
We have demonstrated the evolution of the total energy in simulations and its saturation in the final state. The simulations were also repeated with higher resolution to confirm good convergence. While the stationary solutions studied in this paper are shown to be globally stable and robust, i.e. the rather arbitrary initial states converge to the unique transonic profile, on small scales we observe low amplitude fluctuations of the ion sound type propagating along the nozzle. The amplitude of such fluctuations is larger when the global solutions include the shock wave transitions at the nozzle inlet, as in some cases shown in Figs. \ref{fig:robust} and \ref{Combained fig: rho and M_z space}. A dynamic visualization of the plasma velocity and density evolution is provided in the attached media, illustrating the transition to the transonic regime. This complements the static profiles shown in Figs. \ref{fig:pseudo_plot_v} and \ref{fig:pseudo_plot_rho}. 

\acknowledgments{This work is supported in part by the Natural Sciences and Engineering Research Council of Canada (NSERC). The computational resources were provided by Digital Research Alliance of Canada.}
\section*{Data availability}
The data that support the findings of this study are available from the corresponding author upon reasonable request.
\bibliographystyle{unsrt}
\bibliography{ref}

\begin{thebibliography}{10}

\bibitem{ArefievPoP2004}
A.~V. Arefiev and B.~N. Breizman.
\newblock Theoretical components of the {VASIMR} plasma propulsion concept.
\newblock {\em Physics of Plasmas}, 11(5):2942--2949, 2004.

\bibitem{WuPoP2025}
K.~Wu, Z.~Chen, J.~Ren, Y.~Wang, G.~Zhang, W.~Wang, and H.~Tang.
\newblock A review of plasma acceleration and detachment mechanisms in propulsive magnetic nozzles.
\newblock {\em Physics of Plasmas}, 32(4), 2025.

\bibitem{MerinoPoP2016}
M.~Merino and E.~Ahedo.
\newblock Fully magnetized plasma flow in a magnetic nozzle.
\newblock {\em Physics of Plasmas}, 23(2):023506, 2016.

\bibitem{SchoenbergPoP1998}
K.~F. Schoenberg, R.~A. Gerwin, R.~W. Moses, J.~T. Scheuer, and H.~P. Wagner.
\newblock Magnetohydrodynamic flow physics of magnetically nozzled plasma accelerators with applications to advanced manufacturing.
\newblock {\em Physics of Plasmas}, 5(5):2090--2104, 1998.

\bibitem{RyutovAIP2016}
D.~D. Ryutov, P.~N. Yushmanov, D.~C. Barnes, and S.~V. Putvinski.
\newblock {\em Divertor for a linear fusion device}, volume 1721 of {\em AIP Conference Proceedings}, page 060003.
\newblock 2016.

\bibitem{EndrizziJPP2023}
D.~Endrizzi, J.~K. Anderson, M.~Brown, J.~Egedal, B.~Geiger, R.~W. Harvey, M.~Ialovega, J.~Kirch, E.~Peterson, Yu~V. Petrov, J.~Pizzo, T.~Qian, K.~Sanwalka, O.~Schmitz, J.~Wallace, D.~Yakovlev, M.~Yu, and C.~B. Forest.
\newblock Physics basis for the wisconsin {HTS} axisymmetric mirror {(WHAM)}.
\newblock {\em Journal of Plasma Physics}, 89(5):975890501, 2023.

\bibitem{AndersenPF1969}
S.~A. Andersen.
\newblock Continuous supersonic plasma wind tunnel.
\newblock {\em Physics of Fluids}, 12(3):557, 1969.

\bibitem{KurikiPF1970}
K.~Kuriki and O.~Okada.
\newblock Experimental study of a plasma flow in a magnetic nozzle.
\newblock {\em Physics of Fluids}, 13(9):2262, 1970.

\bibitem{BathgatePSST2017}
S.~N. Bathgate, M.~M. Bilek, and D.~R. McKenzie.
\newblock Electrodeless plasma thrusters for spacecraft: a review.
\newblock {\em Plasma Science \& Technology}, 19(8):083001, 2017.

\bibitem{OnofriPoP2017}
M.~Onofri, P.~Yushmanov, S.~Dettrick, D.~Barnes, K.~Hubbard, and T.~Tajima.
\newblock Magnetohydrodynamic transport characterization of a field reversed configuration.
\newblock {\em Physics of Plasmas}, 24(9):092518, 2017.

\bibitem{Morozov_V8}
A.I. Morozov and L.S. Solovev.
\newblock {\em Steady-State Plasma Flow in a Magnetic Field}, volume~8 of {\em Reviews of Plasma Physics, {\normalfont edited by M.A. Leontovich}}.
\newblock Springer US, New York, 1980.

\bibitem{SteinhauerPoP1999}
L.~C. Steinhauer.
\newblock Formalism for multi-fluid equilibria with flow.
\newblock {\em Physics of Plasmas}, 6(7):2734--2741, 1999.

\bibitem{HameiriPF1983}
E.~Hameiri.
\newblock The equilibrium and stability of rotating plasmas.
\newblock {\em Physics of Fluids}, 26(1):230, 1983.

\bibitem{GuazzottoPoP2005}
L.~Guazzotto and R.~Betti.
\newblock Magnetohydrodynamics equilibria with toroidal and poloidal flow.
\newblock {\em Physics of Plasmas}, 12(5):056107, 2005.

\bibitem{GoedbloedPhysScr2002}
J.~P. Goedbloed.
\newblock Transonic magnetohydrodynamic flows in laboratory and astrophysical plasmas.
\newblock {\em Physica Scripta}, T98:43--47, 2002.

\bibitem{VelliASC2001}
M.~Velli.
\newblock Hydrodynamics of the solar wind expansion - why is the solar wind supersonic?
\newblock {\em Astrophysics and Space Science}, 277(1-2):157--167, 2001.

\bibitem{Tsinganos2007}
K.~Tsinganos.
\newblock {\em Theory of MHD Jets and Outflows}, volume 723 of {\em Lecture Notes in Physics}, pages 117--159.
\newblock Springer Berlin Heidelberg, 2007.

\bibitem{FruchtmanPoP2012}
A.~Fruchtman, K.~Takahashi, C.~Charles, and R.~W. Boswell.
\newblock A magnetic nozzle calculation of the force on a plasma.
\newblock {\em Physics of Plasmas}, 19(3):033507, 2012.

\bibitem{LafleurPoP2014}
T.~Lafleur.
\newblock Helicon plasma thruster discharge model.
\newblock {\em Physics of Plasmas}, 21(4):043507, 2014.

\bibitem{ManheimerIEEE2001}
W.~M. Manheimer and R.~F. Fernsler.
\newblock Plasma acceleration by area expansion.
\newblock {\em {IEEE} Transactions on Plasma Science}, 29(1):75--84, 2001.

\bibitem{SmolyakovPoP2021}
A.~I. Smolyakov, A.~Sabo, P.~Yushmanov, and S.~Putvinskii.
\newblock On quasineutral plasma flow in the magnetic nozzle.
\newblock {\em Physics of Plasmas}, 28(6):060701, 2021.

\bibitem{DubinovJPP2005}
A.~E. Dubinov and I.~D. Dubinova.
\newblock How can one solve exactly some problems in plasma theory.
\newblock {\em Journal of Plasma Physics}, 71(05):715, 2005.

\bibitem{CorlessAdvCompMath1996}
R.~M. Corless, G.~H. Gonnet, D.~E.~G. Hare, D.~J. Jeffrey, and D.~E. Knuth.
\newblock On the {LambertW} function.
\newblock {\em Advances in Computational Mathematics}, 5(1):329--359, 1996.

\bibitem{JimenezPoP2022}
M.~Jimenez, A.~I. Smolyakov, O.~Chapurin, and P.~Yushmanov.
\newblock Ion kinetic effects and instabilities in the plasma flow in the magnetic mirror.
\newblock {\em Physics of Plasmas}, 29(11):112117, 2022.

\bibitem{SaboPoP2022}
A.~Sabo, A.~I. Smolyakov, P.~Yushmanov, and S.~Putvinski.
\newblock Ion temperature effects on plasma flow in the magnetic mirror configuration.
\newblock {\em Physics of Plasmas}, 29(5):052507, 2022.

\bibitem{TyushevPoP2025}
M.~Tyushev, A.~Smolyakov, A.~Sabo, R.~Groenewald, A.~Necas, and P.~Yushmanov.
\newblock Drift-kinetic pic simulations of plasma flow and energy transport in the magnetic mirror configuration.
\newblock {\em Physics of Plasmas}, 32(3), 2025.

\bibitem{DeguireMSc}
J.~Deguire.
\newblock {\em Plasma flow and acceleration in the magnetic mirror and nozzle geometries, University of Saskatchewan, https://hdl.handle.net/10388/16446}.
\newblock M.{S}c. thesis, 2024.

\bibitem{PostNF1987}
R.~F. Post.
\newblock The magnetic mirror approach to fusion.
\newblock {\em Nuclear Fusion}, 27(10):1579--1739, 1987.

\bibitem{Mignone2007}
A.~Mignone, G.~Bodo, S.~Massaglia, T.~Matsakos, O.~Tesileanu, C.~Zanni, and A.~Ferrari.
\newblock {PLUTO:} a numerical code for computational astrophysics.
\newblock {\em The Astrophysical Journal Supplement Series}, 170(1):228--242, 2007.

\bibitem{PiochThesis2024}
R.~Pioch.
\newblock {\em Ion dynamics in the magnetic nozzle of an Electron Cyclotron Resonance Thruster- Dynamiques ioniques dans la tuyère magnétique d'un propulseur ECR, Institut Polytechnique de Paris, https://theses.hal.science/tel-04861648}.
\newblock Thesis, 2025.

\bibitem{SmolyakovPPR2025}
A.~I. Smolyakov, A.~Sabo, S.~Krasheninnikov, and P.~Yushmanov.
\newblock Electromagnetic and centrifugal effects on plasma acceleration in the magnetic nozzle.
\newblock {\em Plasma Physics Reports}, 2025.

\end{thebibliography}
\end{document}